\documentstyle[aps,prl,epsf,floats]{revtex}

\begin{document}

\twocolumn[\hsize\textwidth\columnwidth\hsize\csname@twocolumnfalse\endcsname

\title{Discrete breathers in nonlinear lattices:  Experimental detection in a Josephson array}

\author{E. Tr\'{\i}as$^{1}$, J. J. Mazo$^{1,2}$ and T. P. Orlando$^{1}$}

\address{$^{1}$ Department of Electrical Engineering and Computer Science,\\
Massachusetts Institute of Technology, Cambridge, Massachusetts 02139,\\}
\address{$^{2}$ Departamento de F\'{\i}sica de la Materia Condensada and ICMA \\
CSIC-Universidad de Zaragoza, E-50009 Zaragoza, Spain}

\date{\today}
\maketitle

\tightenlines

\begin{abstract}
We present an experimental study of discrete breathers in an underdamped
Josephson-junction array.  Breathers exist under a range of dc current
biases and temperatures, and are detected by measuring dc voltages.  We find
the maximum allowable bias current for the
breather is proportional to the array depinning
current while the minimum current seems to be related to a junction
retrapping mechanism. We have observed that this latter instability leads to the
formation of multi-site breather states in the array.  We have also 
studied the domain of existence of the breather at different values
of the array parameters by varying the temperature.
\end{abstract}

]

\narrowtext

Discrete breathers are a new type of excitation in nonlinear lattices.
They are characterized by an exponential localization of the energy. This
localization does not occur in linear systems and it is different from
Anderson localization, which is due to the presence of impurities. Thus,
discrete breathers are also known as intrinsic localized modes.

Breathers have been proven to be generic solutions for
the dynamics of nonlinear coupled oscillator systems \cite{mackay94,mackay97}
by the use of the novel mathematical technique of the anti-integrable
limit \cite{aubry90}.
They have been extensively studied \cite{sievers88,sievers95,takeno96,aubry97,flach98}
and have been proposed to theoretically exist in
diverse systems such as in spin wave modes of 
antiferromagnets \cite{sievers98},
DNA denaturation \cite{peyrard98},
and the dynamics of Josephson-junction networks \cite{floria96,mazo99}.
Also, they have been shown to be important in the
dynamics of mechanical engineering systems
\cite{vakakis96,vakakis99}.
Although a number of experiments have been proposed, discrete breathers 
have yet to be experimentally generated and measured.

In this Letter, we present, to our knowledge,
the first experimental study of discrete breathers in a
spatially extended system. 
We have designed and fabricated an underdamped
Josephson-junction ladder which allows for the existence of breathers
when biased by dc external currents. 
We have developed a method for
exciting breathers and explored their existence domain
and instability mechanisms with
respects to the junction parameters and the applied current.

A Josephson junction  consists of  two superconducting leads
separated by a thin insulating barrier.  Due to the  Josephson effect,
it behaves as a solid-state nonlinear oscillator and
is usually modeled by the same dynamical equations that govern
the motion of a driven pendulum\cite{strogatz_book,pendulum}:
$i=\ddot{\varphi} +\Gamma \dot{\varphi} + \sin \varphi$.
The response of the junction to a  current is
measured by the voltage of the junction which is
given by $v = (\Phi_0/2\pi){d \varphi / dt}$.
By coupling junctions it is possible
to construct solid-state physical realizations of different models such
as the Frenkel-Kontorova \cite{watanabe95} model for nonlinear
dynamics and the 2D XY model \cite{resnick81} for phase transitions in
condensed matter.
Moreover, since the parameters,  such as $\Gamma(T)$, vary
with temperature,  a range of parameter space  can be studied easily
with each sample.

\begin{figure}[tb]
\epsfxsize=2.8in
\centering{\mbox{\epsfbox{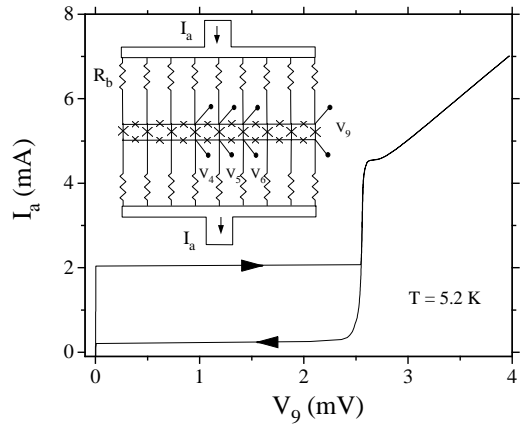}}}
\vspace{0.1in}
\caption[]{
Current-voltage characteristic of an anisotropic Josephson junction array. The
hysteresis between the depinning current
($I_{dep} \approx 2\,{\rm mA}$) and the retrapping current
($I_r \approx 0.2\,{\rm mA}$) is shown.
Inset: Schematic of the anisotropic ladder array.
Vertical junctions have four times the area
of the horizontal ones.
Current is
injected through bias resistors, $R_b$, of $25\,\Omega$.
These resistors are larger than the vertical junction's
normal state resistance of $5 \,\Omega$ to
assure a more homogeneous current drive
and are also larger then the horizontal's normal state
resistance of $20 \,\Omega$ to minimize effects on their
dynamics.
There are voltage probes in the fourth,
fifth, sixth and ninth vertical junctions to measure
$V_4$, $V_5$, $V_6$ and $V_9$.
The  voltage probes can also be used to
measure the top  horizontal junctions in the middle  which we denote as
$V_{4T}$ and $V_{5T}$.
}
\label{fig:rn}
\end{figure}

The inset of Fig.~\ref{fig:rn} shows a schematic
of the anisotropic ladder array.  The  junctions are fabricated using a 
Nb-Al$_2$O$_x$-Nb tri-layer technology with
a critical current density of $1000 \, {\rm A/cm^2}$.
The current is injected and extracted through bias resistors
in order to distribute the current as uniformly as possible through the
array.  These resistors are large enough so as
to minimize any deleterious effects on the dynamics.
The anisotropy of the array is defined by $h$ as
the ratio of areas of the horizontal to
vertical junctions.  In our arrays $h=1/4$ and 
$h=I_{ch}/I_{cv}=R_v/R_h=C_h/C_v$, and $\Gamma_v=\Gamma_h=\Gamma$.
As shown in the schematic,
we have placed voltage probes at various junctions in order to
measure the voltages of both horizontal and vertical junctions.

In Fig.~\ref{fig:rn} we show a typical current-voltage, IV, characteristic
of the array.
As the applied current increases from zero we measure the
average voltage of the 9-th junction.  
The junction
starts at a zero-voltage state and remains there until it reaches the
array's depinning current $I_{dep}$ at about $2\,{\rm mA}$.
The depinning current can roughly be understood as
the sum of the vertical junction's intrinsic critical current $I_{cv}$ and
the small circulating Meissner current around the array.
In a pendulum analogy, the critical current 
is equivalent to the critical torque, which is just 
sufficiently strong enough to force the pendulum to start rotating. 
When the current is larger the junction switches from zero-voltage state
to the junction's superconducting gap voltage, $V_g$, 
which at this temperature is $2.5\,{\rm mV}$.  At this
point all of the vertical junctions are said to be rotating 
and the array is in its ``whirling state''.
One of the effects of this gap voltage
is to substantially affect the junction's
resistance, and thereby damping, in a complicated
nonlinear way.  The current
can be further increased until the junction reaches its normal state and it
behaves as a resistor, $R_n$, of $5\,\Omega$.
As the current decreases
the junction returns to the gap voltage and then
to its zero-voltage state at the 
retrapping  current, $I_r$,
of $\approx 0.2\,{\rm mA}$.
The hysteresis loop between $I_{dep}$ and $I_{r}$ 
is due to our underdamped junctions:
the inertia  causes the junctions to continue to  rotate  when 
the applied current is lowered from above its critical value.

It is this hysteresis loop that allows for the existence of
breathers in the ladder with dc bias current.  In this current range the
zero-voltage ($V=0$) and rotating ($V = V_g$) solutions coexist.  
Then, a
discrete breather in the ladder corresponds to when one vertical junction
is rotating while the other vertical junctions librate. This solution
is easy to conceive in the limit where the vertical junctions are
imagined to be completely
decoupled.  However, whether a localized solution can exist in the ladder
will be determined by
the strength of the spatial coupling between vertical junctions.
This coupling 
occurs through three mechanisms:  flux quantization, 
self and mutual inductances of the
meshes,  and the horizontal junctions.  Though the effective
coupling is a complicated function of the array 
parameters, it is most strongly controlled by $h$.
If the anisotropy $h$ is
too large, then the vertical junctions will not support 
localized solutions that can be excited by dc currents.  
It has been determined from
simulations of the system \cite{mazo99} that
$h=1/4$ will allow for the existence of breathers in
our ladders.

\begin{figure}[tb]
\epsfxsize=2.8in
\centering{\mbox{\epsfbox{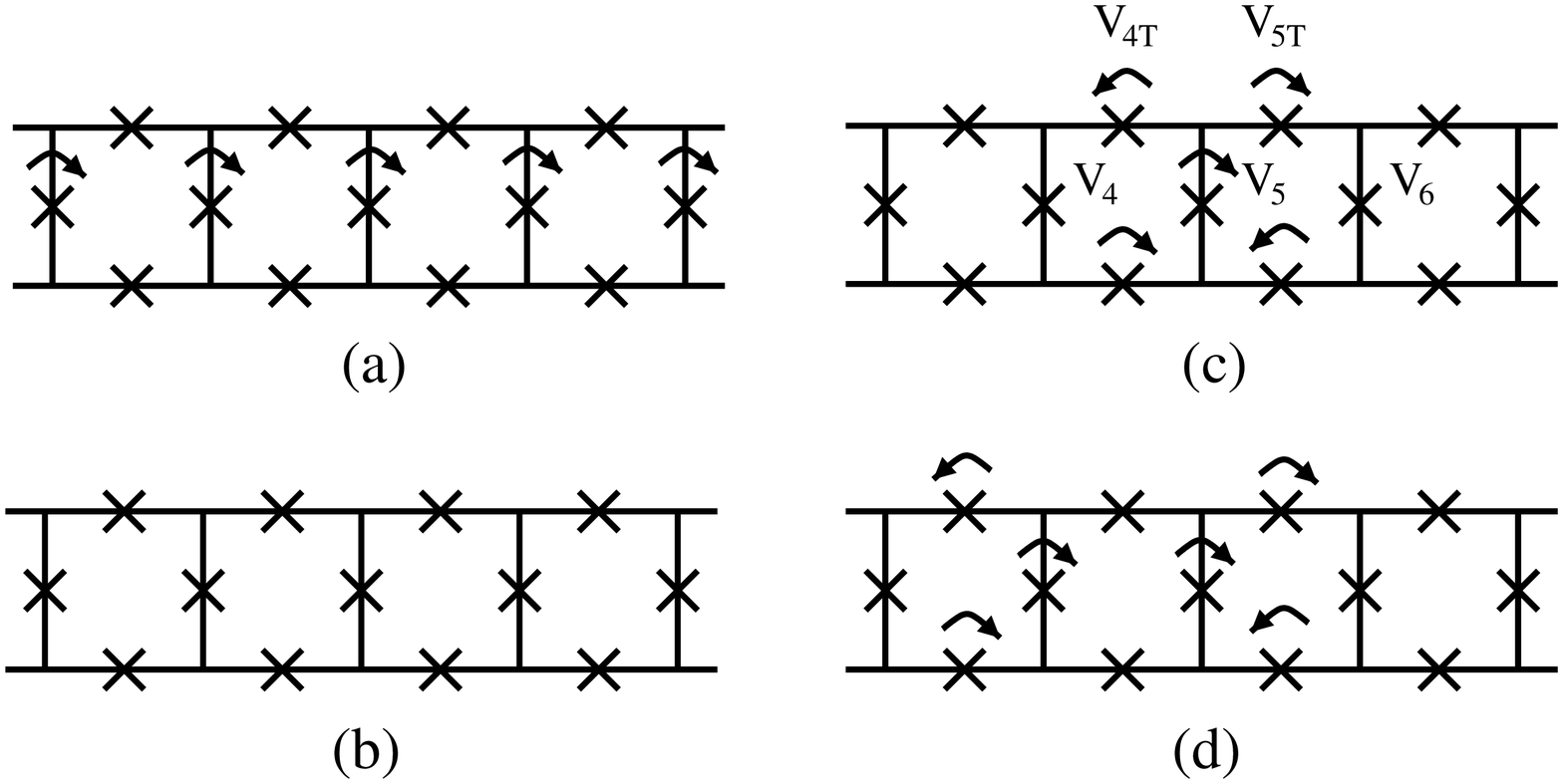}}}
\vspace{0.1in}
\caption[]{Schematic of solutions in the array:
(a) the whirling state where vertical junctions
are rotating and horizontal ones librate;
(b) the zero-voltage state where there are no rotating junctions;
(c) a breather solution where the fifth 
vertical junction and the nearest horizontal neighbors rotate and the other 
vertical and horizontal junctions librate;
(d) a multi-site breather solution where two vertical junctions
rotate.}
\label{fig:arrays}
\end{figure}

Figure~\ref{fig:arrays} shows some possible solutions for the states
of our ladder.  Graph (a) is the whirling state with every
vertical junction rotating as indicated by the arrows.  
This is the state when all the junctions
of the array have switched to the gap voltage.
Graph (b) shows the zero voltage state
with no rotating junctions.  Graph (c) depicts a single-site breather
solution.  Here, 
one of the vertical and the horizontal
neighboring junctions rotate.  The horizontal junctions 
allow the vertical junction to rotate with a mean voltage without
an overall increase of the stored magnetic energy.   Graph (d)
shows a two-sited breather where two vertical junctions
rotate.  We have
experimentally detected these types of
localized solutions (c and d)
by measuring the average dc voltage of the junctions
as labeled in (c).

\begin{figure}[tb]
\epsfxsize=2.8in
\centering{\mbox{\epsfbox{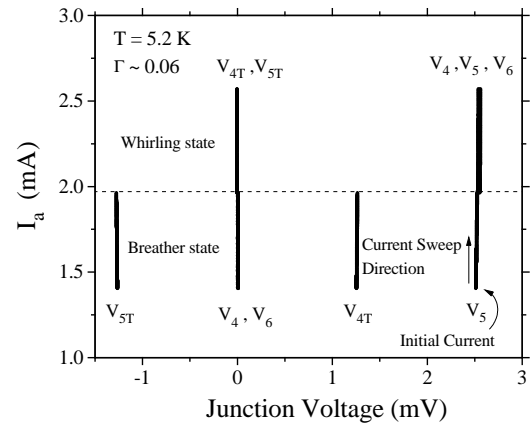}}}
\vspace{0.1in}
\caption[]{Measured averaged voltages of five junctions
in the center of array as the applied current
is increased.  We have biased the ladder at {1.4\,{\rm mA}} 
and excited a breather as indicated in the text.  Then 
the applied current is
increased.  There are two regions in the IV.  Below 
{$\approx 2\,{\rm mA}$} we see the breather; above,
the breather becomes unstable and the array switches to the whirling
state.}
\label{fig:up}
\end{figure}

For our experiments, we have developed a simple 
reproducible method of exciting a breather:
(i) bias the array uniformly to a current below depinning current;
(ii) increase the current injected into the  middle vertical junction 
[labeled $V_5$ in Fig.~\ref{fig:arrays}(c)] until
its voltage switches to the gap;
(iii) reduce this extra current in the middle junction to zero.  

For example, to prepare the initial state in Fig.~\ref{fig:up} we started by increasing
the applied current to {1.4\,{\rm mA}} which is below $I_{dep}$.
At this point the array is in the zero-voltage state.  
We then add an extra bias current to the middle junction (number 5) until 
it switches to the gap voltage of 2.5 mV
and then we reduce this extra bias to zero.  In a sense we have prepared
the initial conditions for the experiment.  We can now increase the uniform
applied current while 
simultaneously measuring the voltages of the vertical junctions ($V_4$, $V_5$ 
and $V_6$)
and the top two horizontal junctions, $V_{4T}$ and $V_{5T}$,
as labeled in Fig.~\ref{fig:arrays}(c).

Figure~\ref{fig:up} shows the result after we have excited the
breather and we have increased the array current.
Close to the initial current of 1.4 mA only the fifth vertical junction
is at $V_g$ and both the fourth and sixth vertical 
junctions are in the zero-voltage
state.  This is the breather state shown in Fig.~\ref{fig:arrays}(c) and 
in essence
the signature of the localized breather: a vertical junction is rotating while
its neighboring vertical junctions do not rotate.
We also see that both neighboring horizontal junctions have a voltage magnitude that is
precisely half of this value ($V_{4T}=-V_{5T}=V_5/2$ and $V_4=V_6=0$).
Both the magnitude and the sign can be
understood by applying
Kirchoff's voltage law
to top-bottom voltage-symmetric solutions, as sketched
in  Fig.~\ref{fig:arrays}(c). The voltage of the top horizontal junction
is equal to the negative of the bottom one.
Since the voltage drops around the loop must be zero,
the horizontal voltages must be half that of the active  vertical
junction voltage.
As we increase the current the breather 
continues to exist until the applied current approaches $I_+ \approx {2\,{\rm mA}}$.
At this point the horizontal junctions switch to a zero-voltage state
while all of the vertical junctions switch to $V_g=2.5$ mV.  The array is now
in its whirling
state as drawn in Fig.~\ref{fig:arrays}(a) where $V_4=V_6=V_5=V_g$ and 
$V_{4T}=V_{5T}=0$.

\begin{figure}[tb]
\epsfxsize=2.8in
\centering{\mbox{\epsfbox{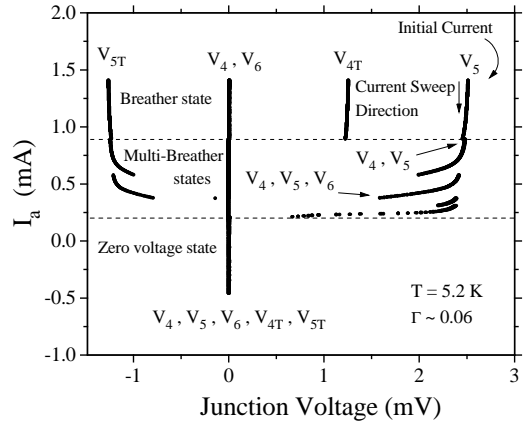}}}
\vspace{0.1in}
\caption[]{Measured average
voltages of five junctions in the center of array as the applied current
is decreased.  We have biased the ladder at {1.4\,{\rm mA}} 
and excited a breather as indicated in the text.  Then
the applied current is decreased.  There are three
regions in the IV.  The breather state
becomes unstable at around  {0.8\,{\rm mA}} 
leading to a sequence of shifts at the gap voltage 
that are interpreted as multi-site breather states.
The array reaches its zero-voltage state at about {0.2\,{\rm mA}}.}
\label{fig:down}
\end{figure}

If we excite the breather again but instead of increasing the applied current 
we decrease it,
we measure curves typical of Fig.~\ref{fig:down}.  As explained above,
we prepare the array in an initial 
condition with a breather located in junction 5 at {1.4\,{\rm mA}}.
We then decrease the applied current slowly.  We start with
the signature measurement of the breather:  junction five is
rotating at $V_g$
while $V_4$ and $V_6=0$.  We also see that the horizontal junctions
have the expected value of $V_g/2$.  As the current is decreased the breather
persists until the array is biased at {0.8\,{\rm mA}}.  The fourth vertical junction 
then switches to the gap voltage while $V_{4T}$ switches to a zero voltage state.  The
resulting array state is sketched in Fig.~\ref{fig:arrays}(d) with
$V_4=V_5=V_g$ while $V_{5T}=V_g/2$ and $V_{4T}=V_6=0$.  The single-site breather
has destabilized by creating a two-site breather.

As the applied current is further decreased beyond
the single-site breather instability at ${\approx 0.8\,{\rm mA}}$,
the voltage of the fourth and fifth vertical junctions decreases but then 
suddenly jumps back to $V_g$.  Then the voltage decreases again, and it again
jumps back to $V_g$.  This second shift corresponds to the sixth junction switching from
the zero voltage state to the gap voltage.
At this current bias, all of the three measured vertical voltages are rotating.  There is
a further jump of the voltage as the current decreases.  Finally, at
{0.2\,{\rm mA}} all of the vertical junctions return to their zero-voltage state
via a retrapping mechanism analogous to that of a single pendulum.  

From these experiments
and corroborating numerical simulations
we speculate that this
shifting of the voltage back to $V_g$ corresponds to at least one
vertical junction switching from the zero-voltage state to the rotating
state.  The shapes of the IV curves in this multi-site breather
regime are influenced by the redistribution of current when each vertical
junction switches.
This redistribution may also govern the evolution
of the system after each transition to one of the other possible 
breather attractors
in the phase space of the array.  However, the exact nature of
the selection process is not yet understood.

\begin{figure}[tb]
\epsfxsize=2.8in
\centering{\mbox{\epsfbox{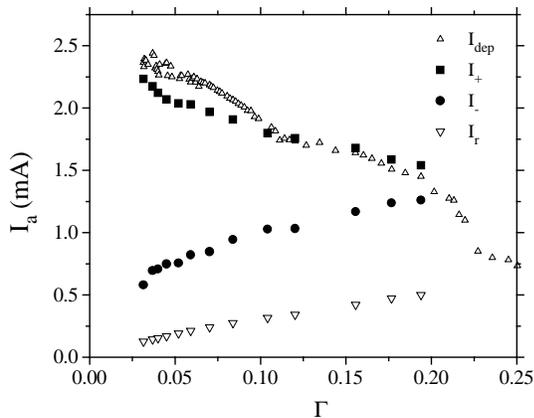}}}
\vspace{0.1in}
\caption[]{Existence region of the breather in 
the current-$\Gamma$ plane.  $\Gamma$ was varied by changing temperature.
$I_{+}$ is the maximum applied current the breather supports,
while $I_{-}$ is the minimum current.  
}
\label{fig:temp}
\end{figure}

The above data was taken at a temperature of 5.2 K.  We
found four current values of importance:  the current
when the array returns to the zero-voltage, $I_r$; the maximum 
zero-voltage state current, $I_{dep}$;
the maximum current the breather supports, $I_{+}$;
and the minimum current $I_{-}$.
By sweeping the temperature we can
study how the current range in which 
our breather exists is affected by a change of
the array parameters.

Figure ~\ref{fig:temp}
shows the results of plotting the four special current values versus $\Gamma$.
To calculate how the junction
parameters vary with temperature we take $I_{cv}(0)R_{n} = 1.9\,\rm mV$
and assume that the junction critical current follows the standard
Ambegaokar-Baratoff dependence \cite{ambegaokar63}.
We estimate $\Gamma$ from $I_r$ by the relation
$I_r/N I_{cv} = (4/\pi)\Gamma$\cite{guckenheimer}, where
$N$ is the number of vertical junctions in the array.  The other
relevant parameter is the dimensionless penetration depth,
$\lambda_{\perp}=\Phi_{0}/2\pi L_s I_{cv}$,
which measures the inductive coupling in the array.
The loop inductance ${L_s}$ is estimated from numerical modeling
of the circuit.
By changing the temperature of the sample, we vary the $I_{cv}$ of
the junction and hence change $\Gamma$ and $\lambda_{\perp}$.
In this sample, the junction parameters
can range from $0.031 < \Gamma < 0.61  $ and $0.04 < \lambda_\perp < 0.43 $
as the temperature varies from 4.2 K to 9.2 K.  
In Fig.~\ref{fig:temp}, $\Gamma < 0.2  $ corresponds
to $T < 6.7 \,{\rm K}$  and $\lambda_\perp < 0.05 $.
At these low
temperatures, there
is a larger variation in $\Gamma$ because of the sensitive dependence
of this parameter to the junction's  resistance below the gap voltage.

As Fig.~\ref{fig:temp} shows,
the maximum current supported by the breather, $I_{+}$, is 
almost equal to $I_{dep}$.  A simple circuit model gives some
physical intuition.
Junctions that are rotating
have some effective resistance while junctions that are in
the zero-voltage state have zero resistance.  In our
breather state, the center junction is rotating.  Therefore,
when we apply a current to our
array the current will tend to flow around the rotating
junction and through the outside junctions
that are in the zero-voltage state.  When
these outside junctions reach their critical currents
they will begin to rotate and the breather
will disappear.  In the simplest
case, when we ignore any circulating Meissner
currents, this model yields $I_+/NI_{cv}=(h+1)/(2h+1)=0.8$.
Since ${I_{dep}}$ is roughly $NI_{cv}$, the depinning current is
the upper bound for the applied
current that the breather can support. 

The instability mechanism that determines $I_{-}$
in our experiments is more difficult to discern.  We offer two
suggestions that are due to the underdamped
character of our system.
One possibility is via
a retrapping mechanism similar to that of a
single junction.  As the middle junction
rotates, it reaches a point where the current
drive is not sufficient to support the rotation
and it destabilizes.  This physical
picture gives $I_{-}/NI_{cv}=(2h+2)(4/\pi)\Gamma$.
So that for our parameters, $I_{-}$ should be
2.5 times larger than $I_r$, as it is
approximately in Fig.~\ref{fig:temp}.
A second possible instability mechanism consist of
resonances between the characteristic frequencies of the breather
and the lattice eigenmodes. 
The breather looses energy as it excites the eigenmodes.
In our experiments,
the breather always looses stability at voltages
close to $V_g$.
For our parameter range, $V_g$ is larger
than the voltages for the lattice eigenmodes,
thus our data seems to favor a retrapping mechanism.
Lastly, we add
that since $I_{dep}$ and
consequently $I_{+}$ decreases with $\Gamma$, there also 
seems to be a critical damping where the breather will cease to exist.
Experimentally we did not find a breather for $\Gamma>0.2$.

In summary, we have experimentally detected different breather and
multi-site breather states in a superconducting Josephson ladder network. 
By varying the external current and temperature
we have studied the domain of existence and the instability
mechanisms of these localized solutions.  In
addition we have also found, but not discussed here, breathers which
are not top-bottom voltage symmetric\cite{mazo99}, in which only the top
(bottom) horizontal junctions rotate while the bottom (top) junctions are 
in the zero-voltage state. These experiments are the first
observations of discrete breathers and multi-site breathers in a 
condensed matter system. 

This work was supported
by NSF grant DMR-9610042 and DGES (PB95-0797).
JJM thanks the Fulbright Commission and the MEC (Spain) for
financial support.
We thank S.~H. Strogatz, A.~E. Duwel, F. Falo, 
L.~M. Flor\'{\i}a, and P.~J. Mart\'{\i}nez
for insightful discussions.


\begin{thebibliography}{10}
\bibitem{pendulum}
The current that passes through the junction $i$ is
normalized by the junction's intrinsic critical current
$I_c(T)$. $\varphi$ is the gauge-invariant phase difference
across the junction and $\Gamma$, the damping, is usually referred to
as the Stewart-McCumber parameter
$\beta_c=\Gamma^{-2}=2\pi I_c C R^2 / \Phi_0$ ($C$ is the capacitance
of the junction, $R$ is its resistance, and $\Phi_0$ is the flux
quantum). Time is normalized by $\tau=\sqrt{ \Phi_0 C / 2\pi I_c}$.
\end{thebibliography}
\end{document}